\documentclass[conference, a4paper]{IEEEtran}
\IEEEoverridecommandlockouts
\usepackage{cite}
\usepackage{amsmath,amssymb,amsfonts}
\usepackage{algorithmic}
\usepackage{graphicx}
\usepackage{textcomp}
\usepackage{xcolor}
\usepackage{multirow}
\usepackage{subfigure}
\usepackage{hyperref}
\usepackage{booktabs}

\def\BibTeX{{\rm B\kern-.05em{\sc i\kern-.025em b}\kern-.08em
    T\kern-.1667em\lower.7ex\hbox{E}\kern-.125emX}}
\begin{document}

\title{Training-Free Continuous Bitrate Control for Scalable Image Coding for Humans and Machines}

\author{\IEEEauthorblockN{Yui Tatsumi}
\IEEEauthorblockA{\textit{Graduate School of FSE,} \\
\textit{Waseda University}\\
Tokyo, Japan \\
yui.t@fuji.waseda.jp}
\and
\IEEEauthorblockN{Hiroshi Watanabe}
\IEEEauthorblockA{\textit{Graduate School of FSE,} \\
\textit{Waseda University}\\
Tokyo, Japan \\
hiroshi.watanabe@waseda.jp}
}

\maketitle

\begin{abstract}
Continuous variable-rate compression is highly demanded in real-world applications, but remains underexplored in scalable image coding for humans and machines.
In this paper, we propose a training-free variable-rate scalable image coding framework.
By adaptively adjusting quantization step sizes based on predicted scale values, the proposed method enables independent and continuous bitrate control for the machine and enhancement layers while preserving important latent information in each layer.
Experimental results demonstrate the effectiveness of the proposed method and highlight the importance of bitrate allocation between the two layers.
\end{abstract}

\section{Introduction}
Learned image compression (LIC) has been extended beyond human-oriented reconstruction to image coding for machines (ICM).
Scalable image coding for humans and machines further supports both objectives within a single architecture, making it suitable for applications such as traffic monitoring, where images are mainly analyzed by recognition models and only occasionally viewed by humans.

However, most existing LIC-based scalable coding methods do not support continuous bitrate control, which limits their practical applicability.
For two-layer scalable coding structures designed for recognition and reconstruction, a dedicated variable-rate method is required to control bitrate allocation between the two dependent layers.

In this paper, we propose a training-free variable-rate scalable image coding framework. 
By extending scale-aware adaptive quantization control to each layer, the proposed method enables independent and continuous bitrate control for humans and machines.
We also experimentally highlight the bitrate allocation problem between these layers.
The training-free and flexible control design makes the framework practical for deployment under diverse real-world situations.

\section{Related Work}

\begin{figure}[t]
\centering
\includegraphics[width=\linewidth]{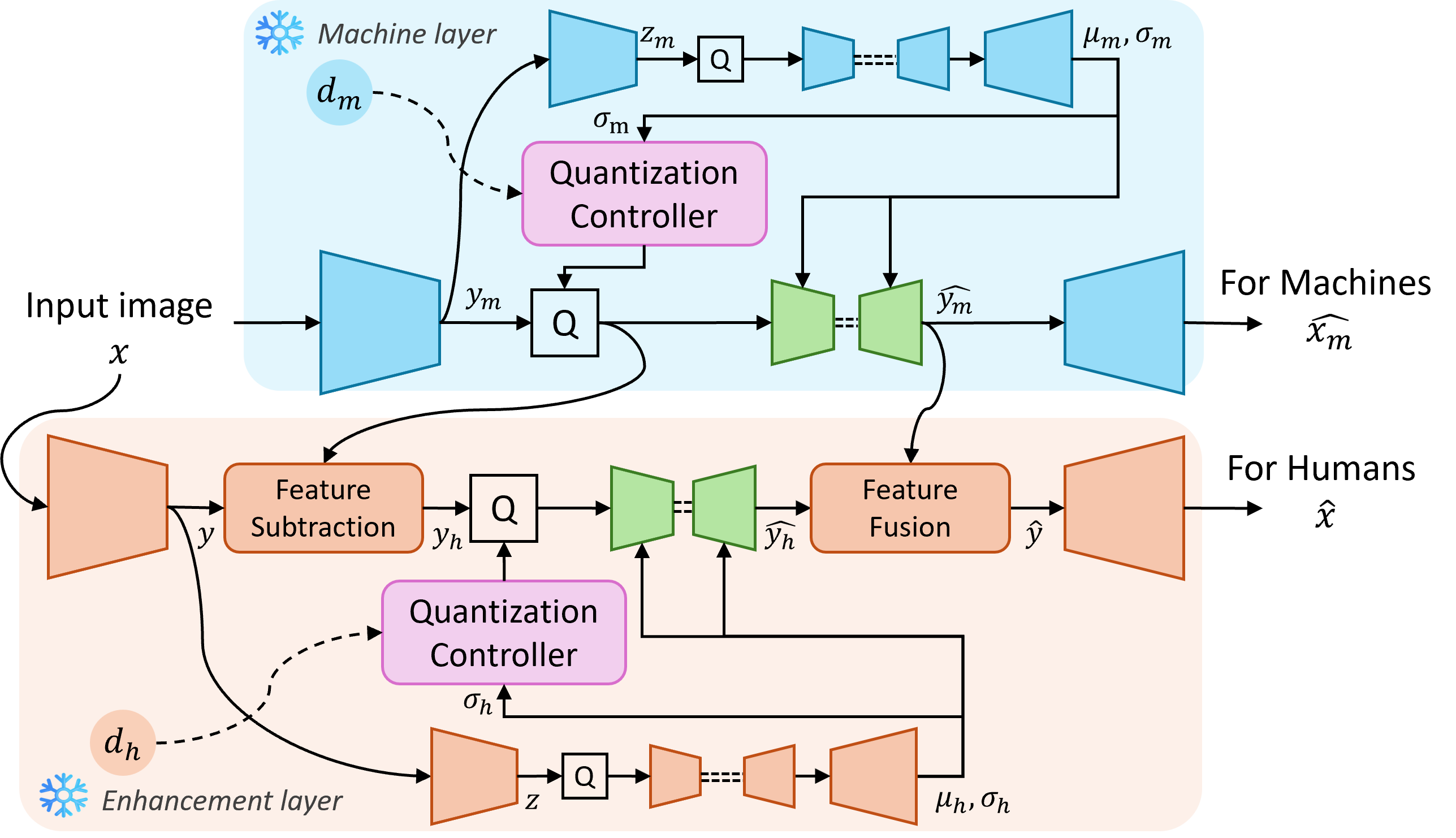}
\caption{Overall architecture of the proposed framework.}
\label{fig:model}
\end{figure}

\subsection{Scalable Image Coding for Humans and Machines}
FR-ICMH \cite{MMSP2025} is a two-layer scalable coding method that employs the task-agnostic SA-ICM \cite{SA-ICM} in the machine layer and compresses explicit feature-level residual information to reconstruct human-oriented images in the enhancement layer. Though effective, it supports only fixed-rate operation.

\subsection{Variable-Rate Image Coding for Machines}
AQVR-ICM~\cite{AQVR-ICM} enables training-free continuous bitrate control for ICM through scale-aware quantization. 
Instead of uniformly increasing quantization steps for all latent elements to reduce bpp, it assigns finer quantization to high-scale recognition-relevant components and coarser quantization to others.
This selective protection maintains higher recognition accuracy under the same bpp reduction.


\section{Proposed Method}
We propose a training-free variable-rate scalable coding framework that extends the scale-aware quantization controller of AQVR-ICM to the two-layer FR-ICMH structure, as shown in Fig.~\ref{fig:model}.
Since the enhancement layer encodes residual information, its latents may also exhibit non-uniform importance.

The machine and enhancement layers are independently and continuously controlled by user-given parameters $d_m$ and $d_h$, respectively.
For simplicity, we use $d$ and $\sigma$ to denote the control parameter and predicted scale parameter of either layer in the following formulation.
Since FR-ICMH employs a channel-wise autoregressive entropy model \cite{ch-arm}, the latent representation in each layer is divided into $N$ channel slices.

The controller determines quantization step sizes in two stages.
First, it sets slice-wise lower and upper bounds so that earlier slices, which contain more important features, are quantized more finely.
For the $n$-th slice, these bounds are:
\begin{equation}
\Delta_{\min}^{(n)} =
\begin{cases}
d + \frac{n-1}{N}(1-d), & 0<d<1,\\
1, & d \ge 1,
\end{cases}
\end{equation}
\begin{equation}
\Delta_{\max}^{(n)} =
\begin{cases}
1, & 0<d\le 1,\\
1 + \frac{n}{N}(d-1), & d>1.
\end{cases}
\end{equation}

Second, within each slice, the controller maps the predicted scale values to the slice-wise range.
Let $\sigma_{c,h,w}$ denote the scale value at channel $c$ and spatial position $(h,w)$.
For channel $c$ belonging to the $n$-th slice, we compute
\begin{equation}
\sigma_{\max}^{(c)}=\max_{h,w}\sigma_{c,h,w}, \qquad
\sigma_{\min}^{(c)}=\min_{h,w}\sigma_{c,h,w}.
\end{equation}
The adaptive quantization step size is then given by
\begin{equation}
\Delta_{c,h,w}
=
\Delta_{\max}^{(n)}
-
\frac{\sigma_{c,h,w}-\sigma_{\min}^{(c)}}
{\sigma_{\max}^{(c)}-\sigma_{\min}^{(c)}+\epsilon}
\left(
\Delta_{\max}^{(n)}-\Delta_{\min}^{(n)}
\right).
\end{equation}

Since the scale parameters can be obtained through the hyperprior network during inference, the controller can be applied to pretrained scalable coding models without training.

\begin{table}[t]
    \centering
    \caption{mAP@50:95 (\%) comparison with different $d_m$ values}
    \label{tab:aqvr_icm_recognition}
    \begin{tabular}{ccccc}
        \hline
        \multirow{2}{*}{$d_m$} 
        & \multirow{2}{*}{bpp} 
        & \multicolumn{2}{c}{Detection} 
        & Segmentation \\
        \cmidrule(r){3-4}\cmidrule(l){5-5}
        & & YOLOv5 & Mask R-CNN & Mask R-CNN \\
        \hline
        1  & 0.227 & 41.9 & 35.6 & 32.1 \\
        2  & 0.186 & 41.0 & 34.5 & 31.2 \\
        4  & 0.143 & 39.5 & 32.5 & 29.2 \\
        8  & 0.103 & 35.6 & 28.5 & 25.5 \\
        16 & 0.069 & 29.0 & 21.8 & 19.1 \\
        \hline
    \end{tabular}
\end{table}

\begin{figure}[t]
\centering
\includegraphics[width=\linewidth]{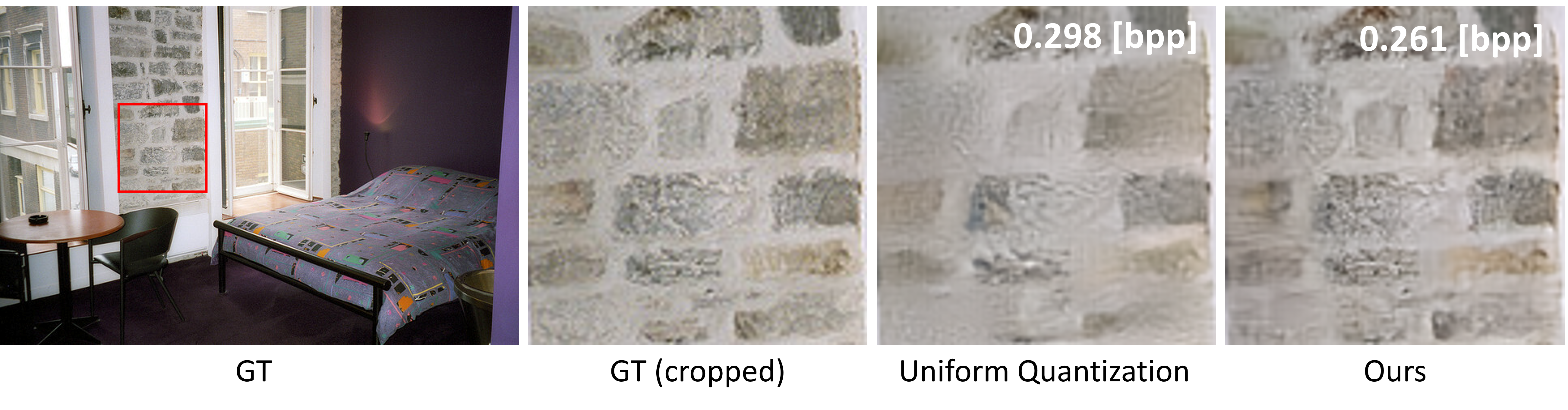}
\caption{Qualitative comparison of human-oriented reconstructed images.}
\label{output_images}
\end{figure}

\section{Experiment}
The COCO-val dataset \cite{COCO} is used for evaluation. We apply the proposed bitrate control to the highest-bitrate pretrained FR-ICMH model without retraining. Table~\ref{tab:aqvr_icm_recognition} shows recognition performance for different $d_m$ values. 
As $d_m$ increases, the bitrate decreases with a gradual accuracy drop. These results are consistent with AQVR-ICM and confirm variable-rate machine-layer control within the scalable coding structure.

\begin{figure}[t]
\centering
\includegraphics[width=\linewidth]{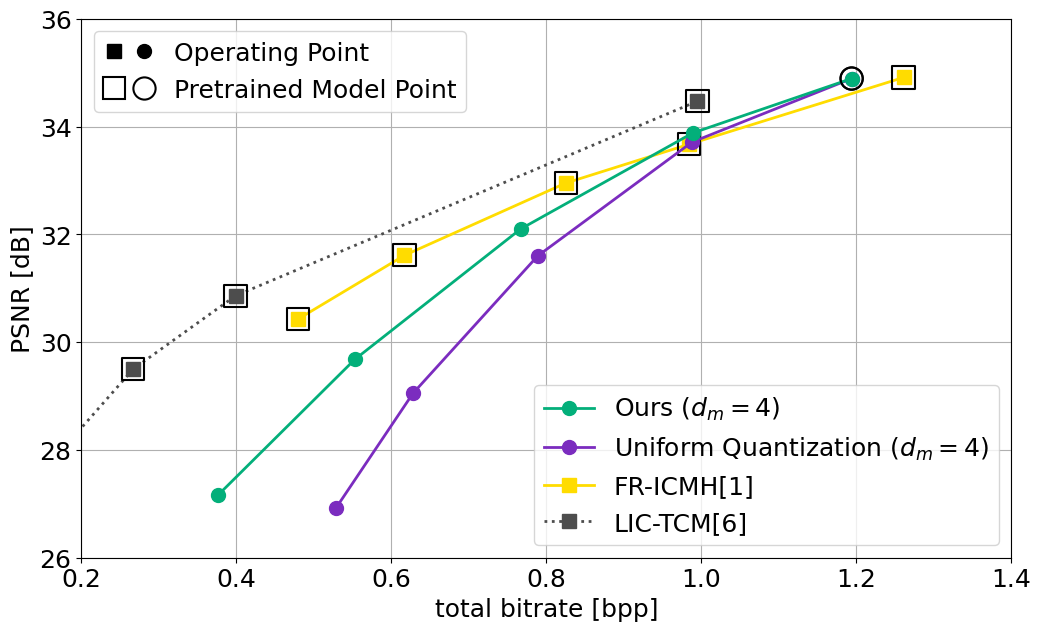}
\caption{Rate-distortion comparison of image reconstruction for human viewing among different methods. Solid/dotted lines indicate scalable/human-oriented codecs, and circle/square markers indicate variable-rate/fixed-rate results.}

\label{graph:method comparison}
\end{figure}

Fig.~\ref{output_images} shows reconstructed examples, where the proposed method preserves wall textures more clearly than uniform quantization at a lower bpp.
Fig.~\ref{graph:method comparison} compares human-oriented compression performance of the proposed method with uniform quantization control, fixed-rate scalable method FR-ICMH, and human-oriented LIC-TCM \cite{LIC-TCM}.
We fix $d_m=4$ and vary $d_h \in \{1,2,4,8,16\}$ for the scalable codecs.
The proposed method outperforms uniform quantization, demonstrating the effectiveness of scale-aware adaptive quantization.

\begin{figure}
\centering
\subfigure[Enhancement-layer bitrate]{%
\includegraphics[clip, width=0.48\hsize]{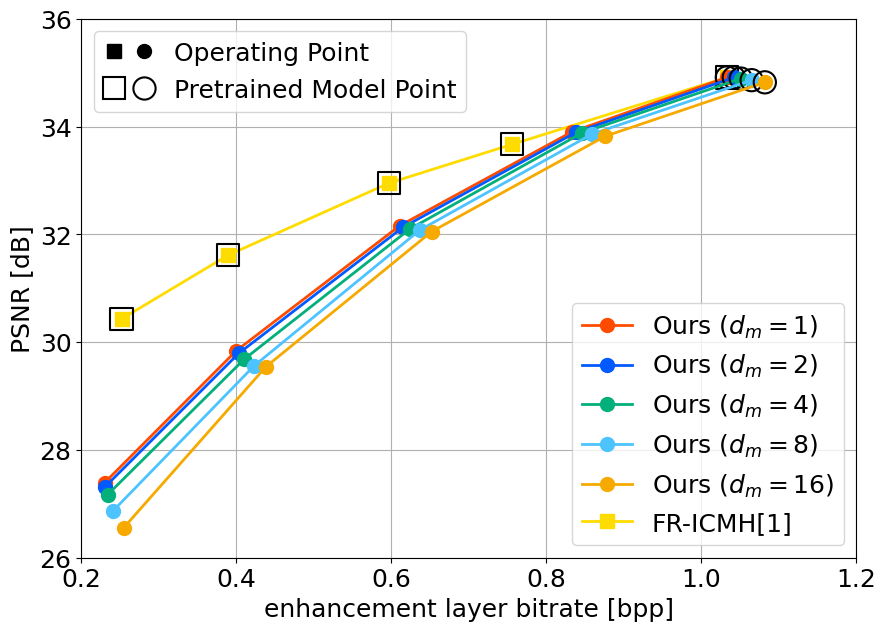}}%
\subfigure[Total bitrate]{%
\includegraphics[clip, width=0.48\hsize]{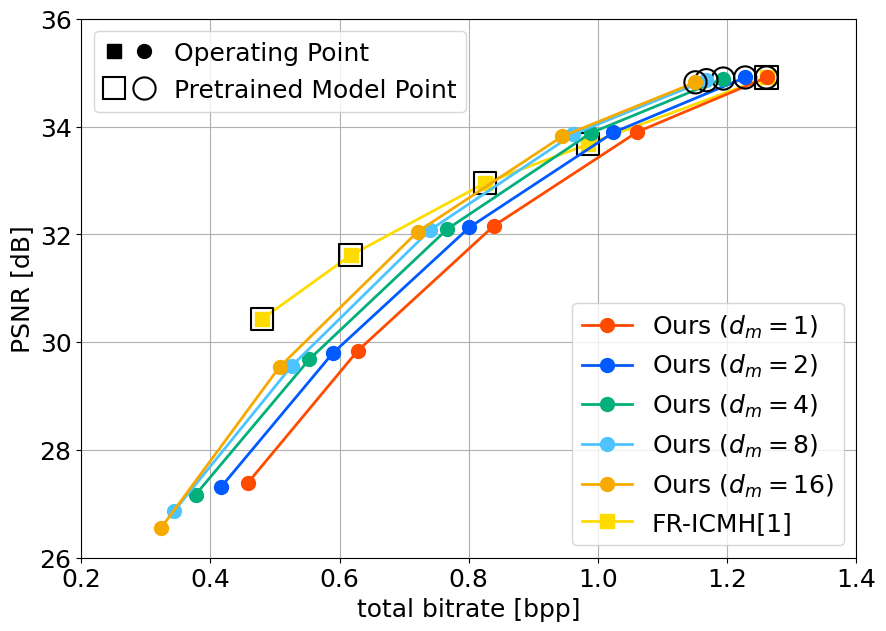}}%
\caption{Effect of $d_m$ on image reconstruction quality in the proposed method with (a) enhancement-layer and (b) total bitrate.}
\label{fig:dm_comparison}
\end{figure}

We then analyze the effect of different machine-layer operating points on human-oriented reconstruction quality.
Fig.~\ref{fig:dm_comparison}(a) and Fig.~\ref{fig:dm_comparison}(b) plot the enhancement-layer and total (machine + enhancement layer) bitrates, respectively.
In terms of the enhancement-layer bitrate, smaller $d_m$ achieves higher PSNR because the enhancement layer receives a more accurate machine-layer representation. 
In contrast, with the total bitrate, larger $d_m$ can provide a better RD trade-off by reducing the machine-layer bitrate. 
This indicates that the optimal $d_m$ depends on the priority between recognition and total human-oriented RD performance, suggesting the need for careful two-layer bitrate allocation in practical applications.

\section{Conclusion}
In this paper, we propose a training-free continuous bitrate control method for scalable image compression for humans and machines. 
Experimental results demonstrate that the proposed method outperforms the baseline and that the preferred bitrate allocation depends on the target objective. 
Future work will investigate automatic bitrate allocation between the layers.

\end{document}